\documentclass[pre,twocolumn,showpacs,superscriptaddress,floatfix]{revtex4}
\usepackage{graphicx}
\usepackage{float}

\begin{document}

\title{Transmission properties of a single metallic slit: From the subwavelength
regime to the geometrical-optics limit}
\author{J. Bravo-Abad}
\affiliation{\mbox{Departamento de F\'{\i}sica Te\'{o}rica de la Materia Condensada
, Universidad Aut\'onoma de Madrid, E-28049 Madrid, Spain}}
\author{L. Mart\'{\i}n-Moreno}
\affiliation{\mbox{Departamento de F\'{\i}sica de la Materia Condensada, ICMA-CSIC,
 Universidad de Zaragoza, E-50009 Zaragoza, Spain}}
\author{F.J. Garc\'{\i}a-Vidal}
\affiliation{\mbox{Departamento de F\'{\i}sica Te\'{o}rica de la Materia Condensada
, Universidad Aut\'onoma de Madrid, E-28049 Madrid, Spain}}

\begin{abstract}

In this work we explore the transmission properties of a single slit in a metallic
screen. We analyze the dependence of these properties on both slit width 
and angle of incident radiation. We study in detail the crossover between the
subwavelength regime and the geometrical-optics limit. In the subwavelength regime,
resonant transmission linked to the excitation of waveguide resonances is analyzed.
Linewidth of these resonances and their associated electric field intensities are 
controlled by just the width of the slit. More complex transmission spectra appear
when the wavelength of light is comparable to the slit width. Rapid oscillations 
associated to the emergence of different propagating modes inside the slit are the 
main features appearing in this regime. 

\end{abstract}

\pacs{78.66.Bz, 42.25.Bs, 42.25.Fx}

\maketitle


\section{Introduction}
After the appearance of the experiment of Ebbesen {\it et al.}
\cite{Ebbesen} showing extraordinary optical transmission through
a 2D array of subwavelength holes perforated in a thick metallic
film, there has been a renewed interest in analyzing the optical
properties of its 1D analog, an array of subwavelength slits
\cite{Schroter,Treacy,Porto,Astilean,Sambles1,Collin,Cao,Barbara,Garcia-Vidal1}.
More recently, resonant transmission  properties of a single subwavelength
slit in a thick metallic screen have been analyzed both
theoretically \cite{Takakura} and experimentally
\cite{Sambles2,Sambles3,Sambles4}. However,
all these works consider the subwavelength regime. To
the best of our knowledge, the study of the crossover between this
regime and the geometrical-optics limit is still lacking.

In this work, we explore how the contribution to the transmission
current of the different modes inside a single slit evolves as a
function of the incident angle and the slit width. To get physical
insight, we apply a formalism based on a modal
expansion of the electromagnetic fields already proposed in an
equivalent problem \cite{Martin-Moreno,Garcia-Vidal2}. As we are
interested in overall transmission behavior rather than in precise
numerical comparison with experiments, we assume infinite conductivity 
for the metallic regions. Note, however, this approximation provides
results of at least semi-quantitative value for good metals, such as silver or gold, 
if the dimensions of the structure are several times larger than the skin 
depth\cite{Garcia-Vidal1}. A bonus of using perfect metal boundary 
conditions is that all theoretical results are directly exportable 
to other frequency regimes, simply by scaling appropriately the 
geometrical parameters defining the structure.

This paper is organized as follows. In Sec. II we describe the
theoretical formalism. This approach is applied to the
subwavelength regime in Sec. III, where a comparison with
available experimental data is also given. In addition, Sec. III
presents two applications of a single subwavelength slit
based on its characteristic resonant behavior. In Sec. IV we analyze the
crossover between the subwavelength and the geometrical-optics
limits. Finally, in Sec. V we summarize our results.

\section{Theoretical framework}

In this work we are interested in the scattering process between a plane
wave and the structure sketched in Fig. 1, i.e., a single metallic
slit in a thick metal screen. Due to the translational
invariance of the system along the $y$ axis, it is possible to write
the vectorial Maxwell equations as two decoupled differential equations,
one of them corresponding to s-polarization (electric field, ${\bf E}$, 
parallel to $y$ axis) and another corresponding to p-polarization
(magnetic field, ${\bf H}$, parallel to $y$ axis). The case of s-polarization
has been throughly analyzed in \cite{Schouten}. Here, we concentrate on 
p-polarized light in which resonant transmission properties have been
reported. For this polarization, the response of
the structure can be solved by finding the solutions of the following equation

\begin{equation}
\nabla \left[ {\frac{1}{\epsilon ({\bf r})}}\nabla  H_y ({\bf r}) \right] +{\frac{\omega ^{2}}{%
c^{2}}} H_y ({\bf r})=0
\end{equation}
where we have assumed a harmonic time dependence of angular frequency
$\omega$. $H_y ({\bf r})$ is the y-component of the magnetic field and 
${\bf r}=(x,z)$. $\epsilon ({\bf r})$ defines the spatial distribution of
the dielectric constant. 

Once we know the field $H_{y} ({\bf r})$, the spatial components of the 
electric field can be obtained from 

\begin{eqnarray}
 E_{x} &=& -{\frac{i}{k_0 \epsilon ({\bf r})}} \frac{\partial H_y}{\partial z} \; ,{\rm and},\\ 
 E_{z} &=& \frac{i}{k_0 \epsilon ({\bf r})} \frac{\partial H_y}{\partial x}    
\end{eqnarray}

To solve Eq. (1) it is convenient to divide the space into three different
regions (labelled as I,II and III in Fig. 1). The magnetic field $H_y$ can
be expressed in regions I and III as a superposition of propagating and
evanescent plane waves as

\begin{eqnarray}
H_y^I({\bf r})&=&H_y^{inc}({\bf r})+ \int_{-\infty}^{\infty} {\rm d}k \: \rho (k) %
\: \exp[i(kx-k_z z)] \\ 
H_y^{III}({\bf r})&=& \int_{-\infty}^{\infty} {\rm d}k \: \tau (k) %
\: \exp[i(kx+k_z z)]
\end{eqnarray}
where $k_z=\sqrt{k_0^2-k^2}$, being $k_0$ the modulus of the wavevector of
the incident plane wave. The coefficients $\rho (k)$ and $\tau (k)$ are
the transmission and reflection amplitudes, respectively.
The incident field is given by 
$H_y^{inc}({\bf r})= \exp[i(k_{0x} x+ k_{0z} z)]$, being
$k_{0x}$ and $k_{0z}$ the projections of the incident wavevector on the x 
and z axis, respectively (see Fig. 1).

\begin{figure}
\begin{center}
\includegraphics[width=8.5 cm]{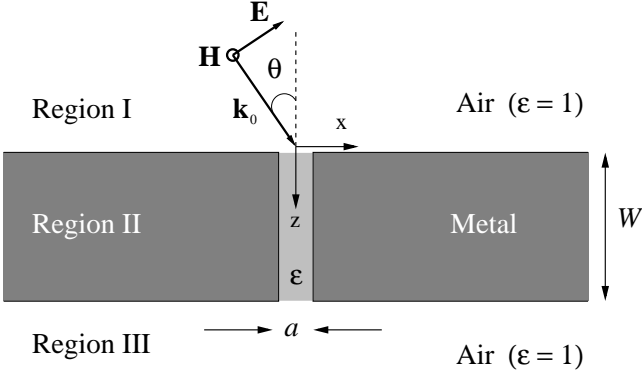}
\end{center}
\caption{Schematic plot of the structure under study. Dark grey corresponds to
metallic regions while light gray corresponds to a region defined by a 
homogeneous dielectric constant $\epsilon$. Both the reference system and
the geometrical parameters that define this system are plotted.}
\end{figure}
 
Inside the slit, the solution can be expanded as a sum of the eigenmodes
of a waveguide

\begin{equation}
H_y^{II} ({\bf r}) = \sum_{m=0}^{\infty} \: [A_{m} \: \exp (i \beta_{m} z)  +
\: B_{m} (-i \beta_{m} z) ]\: \phi_{m} (x)
\end{equation}
where \mbox{$\beta_{m}=[k_0^2 \epsilon -(m\pi/a)^2]^\frac{1}{2}$} and 
the function \mbox{$\phi_{m}(x)=(2/\sqrt{a}) \cos[m\pi/a(x+a/2)]$} 
corresponds to the solution of Eq. (1) within the slit, assuming
perfect metal boundary conditions at $x= \pm a/2$. A homogeneous material
without absorption and characterized by a dielectric constant $\epsilon$, is 
considered inside the slit (light gray area in Fig. 1). By allowing
$\epsilon$ to vary we plan to check the reliability of a single slit
as a fluidics detector. 

By applying the corresponding matching conditions to both $H_y$ and $E_x$
at $z=0$ and $z=W$ and proyecting the resulting equations 
on waveguide modes and plane waves, we obtain a system of linear equations with the set of
coefficients $\{ A_m,B_m \}$ as unknowns. We find that this problem is
equivalent to that reported very recently in Refs. 15 and 16. In
other words, the scattering problem studied in this paper is equivalent to
consider a finite set of indentations with only one relevant mode inside
each indentation. Therefore, to obtain more physical understanding of the mechanisms 
that take part in this scattering problem, it is convenient to change the set 
of variables $\{ A_m,B_m \}$ by the new set $\{E_m^I,E_m^O \}$ defined by

\begin{eqnarray}
E_{m}^{I}&=&\frac{\beta_{m}}{k_0 \epsilon } (A_{m}-B_{m}) \\
E_{m}^{O}&=&\frac{\beta_{m}}{k_0 \epsilon }[A_{m} \exp{(i\beta_{m}W)}-B_{m}\exp{(-i\beta_{m}W)}] 
\end{eqnarray}

The set of equations to solve for the amplitudes
$\{E_m^{I},E_m^{O}\}$ can be expressed as
\begin{eqnarray}
(f_{m} + g_{mm}) E_{m}^{I}+\sum_{m'\ne m } g_{mm'}E_{m'}^{I}+ g_{m}^{v} E_{m}^{O}&=&I_{m}\\
(f_{m} + g_{mm}) E_{m}^{O}+\sum_{m''\ne m} g_{mm''} E_{m''}^{O}+ g_{m}^{v} E_{m}^{I}&=&0
\end{eqnarray}

We can interpret the equations (9) and (10) as follows. The
$m$-th mode has two associated field amplitudes, one of them
($E_{m}^{I}$) corresponds to the scattering events that take place
at the input surface of the slit, while the another amplitude
($E_{m}^{O}$) is related to the output surface. Each amplitude
is determined as the result of different scattering processes 
between all modes. Now, we pass to analyze in detail the different
terms appearing in the set of equations (9) and (10). 

$I_m$ is the overlap integral between the incident plane wave and
$m$-th mode

\begin{equation}
I_{m} = 2 \int_{\frac{-a}{2}}^{\frac{a}{2}} \textrm{dx} \: \phi_{m}(x) \: \exp (ik_{0}\sin \theta \: x) 
\end{equation}

Modes inside the slit are coupled through $g_{mm'}$

\begin{equation}
g_{mm'}=\int_{\frac{-a}{2}}^{\frac{a}{2}} \int_{\frac{-a}{2}}^{\frac{a}{2}} \textrm{dx} \: \textrm{dx'} 
\phi_{m}^{*} (x) \:  \phi_{m'} (x')\:G({\bf {\hat x}  r} , {\bf {\hat x'} r'})
\end{equation}

The function $G({\bf r },{\bf r'})=(\pi/\lambda)
H_{0}^{(1)}(k_{0}|{\bf r} -{\bf r'}|)$ is the two-dimensional Green's 
function corresponding to the vacuum, being
$H_{0}^{(1)}(x)$ the Hankel function of the first kind.
The unitary vectors ${\bf {\hat x}}$ and ${\bf {\hat x'}}$ correspond to
{\bf x} and {\bf x'} axis, respectively. Notice that the propagator $g_{mm'}$ 
does not depend on the dielectric constant inside the slit but on the dielectric
constant of the material outside, i.e., the coupling between modes takes 
places through the external medium.

Amplitude of the $m$-th mode is inversely proportional to $f_{m}+g_{mm}$, where
$f_m$ is the associated admittance of that mode and $g_{mm}$ is the self-interaction term. 

\begin{equation}
f_{m}=  k_{0} \: \epsilon \: \gamma_{m} \frac {\exp(-i\beta_{m}W)+\exp(i\beta_{m}W)}{\beta_{m}[\exp(-i\beta_{m}W) %
-\exp(i\beta_{m}W)]} 
\end{equation}
where $\gamma_{m}$ comes from the normalization factors of the wavefield inside
the slit and it is equal to 1 if $m=0$ and 1/2 if $m \ne 0$.

Amplitude of the $m$-th mode corresponding to the input surface is coupled to its counterpart 
in the output side through $g_{m}^{v}$

\begin{equation}
g_{m}^{v}=  k_{0} \: \epsilon \: \gamma_{m} \frac {2}{\beta_{m} (\exp[i\beta_{m}W)-\exp(-i\beta_{m}W)]}
\end{equation}

It can be proved that in the case of a non uniform slit, modes with different $m$ values are also 
coupled.

It is possible to write the Fourier coefficients $\rho (k)$ and
$\tau (k)$ as a function of $\{E_m^I,E_m^O \}$ 

\begin{eqnarray}
\rho (k) & = & \delta (k-k_{0x}) - \sum_{m=0}^{\infty} E_m^I \: S_m (k) \:\: \textrm{and} \\
\tau (k) & = & \sum_{m=0}^{\infty} E_m^O \: S_m (k)
\end{eqnarray}
where

\begin{equation}
S_m (k) = \frac{1}{2\pi \sqrt{1-(k/k_0)^2}} \int_{- \frac{a}{2}}^{ \frac{a}{2}} \textrm{dx} \: %
\phi_{m} (x)  \: \exp (-ikx) 
\end{equation}

Therefore the field $H_y$ in air regions can be obtained from 
 
\begin{eqnarray}
H_{y}^{I} ({\bf r}) & = & H_{y}^{0} ({\bf r}) +\sum_{m=0}^{\infty} (-E_{m}^{I}) \: G_{m} ({\bf r}) \\
H_{y}^{III} ({\bf r}) & = &\sum_{m=0}^{\infty} E_{m}^{O} \: G_{m} ({\bf r})  
\end{eqnarray}
where we define $H_{y}^{0} ({\bf r}) =2 \cos (k_{0z}z) \exp (ik_{0x}x)$ and

\begin{equation}
G_{m} ({\bf r})= \int_{-\frac{a}{2}}^{\frac{a}{2}} \textrm{dx'} \: \phi_{m} (x')  \: G(\bf{r},\bf{r'})
\end{equation}
with $\bf{r}$=(x,z) and being ${\bf r'}=(x',0)$ for the reflection
region ($z<0$) and ${\bf{r'}}=(x',W)$ for the transmission region
($z>W$). 

Equations (18) and (19) have a clear physical interpretation: scattered fields 
outside the sample are the result of the
sum of those corresponding to a flat dielectric-metal interface
$[H_{y}^{0}({\bf r})]$ and the contribution from each mode within
the slit considered as an individual scatterer.

We can define the transmission current in the propagation direction as

\begin{equation}
J_{z}=\int_{-\infty}^{\infty} \textrm{dx} \: \Re \{[{\bf E^{*}(r)} \times {\bf H (r)}]_{z}\}
\end{equation}

If we normalize the transmission of the system ($T$) to the
flux of energy that impinges on the slit we can write for region III

\begin{equation}
T =\frac{1}{a} \: \Re (\sum_{m,m'=0}^{\infty} E_{m'}^{O} E_{m}^{O*} \: g_{mm'})
\end{equation}

As no absorption is present in the structure, current is conserved
and the transmittance can be then alternatively and more easily 
evaluated inside the slit (region II)

\begin{equation}
T =\frac{1}{a} \: \Re (\sum_{m}^{\infty} E_{m}^{I*} E_{m}^{O} \: g_{m}^v)
\end{equation}

\section{Transmission in the subwavelength regime}

First of all, we consider the transmission properties in the
subwavelength limit, i.e., when $\lambda/a \gg 1$. In order to
compare our numerical results with available experimental data,
we begin by asumming $W=28.2$ mm and the same range of wavelengths
considered in Ref. 12.

\begin{figure}
\begin{center}
\includegraphics[width=7 cm]{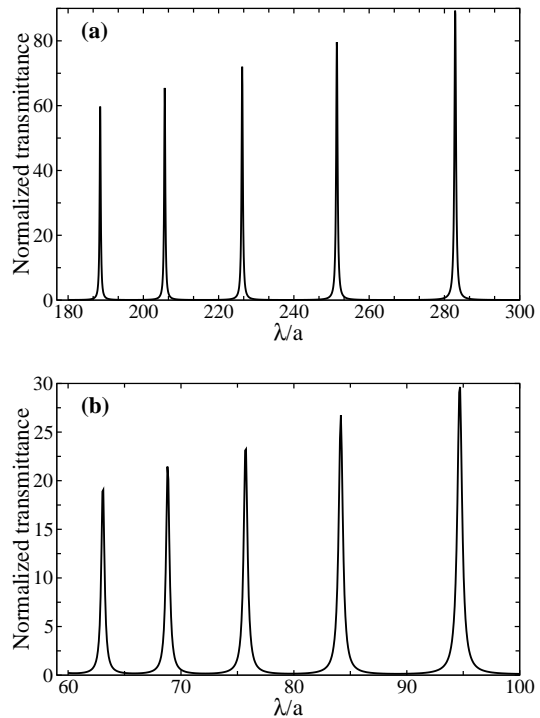}
\end{center}
\caption{Transmittance of a metallic single slit as a function of
the wavelength of the incident radiation, computed with the same
W=28.2 mm and two different values of $a$, 25 $\mu$m (a) and 75
$\mu$m (b). Notice that the wavelength range considered is the same 
in the two panels. Normal incidence and $\epsilon$=1 inside the slit
is assumed.}
\end{figure}

Figure 2 shows the transmittance in two different ranges of
$\lambda/a$ within the subwavelength regime [in Fig. 2(a) we take 
$a$=25 $\mu$m while in Fig. 2(b) we consider $a$=75 $\mu$m]. 
In this figure we assumed normal incidence and $\epsilon$=1 within the slit. 
In the subwavelength regime, only the propagating 
$m=0$ mode has a non-negligible contribution to the transmittance.
Then, for this particular case, the set of equations (9)-(10)
converts into a system of two linear equations with a solution
for $E_{0}^{O}$

\begin{equation}
E_{0}^{O}=-\frac{g_0^v I_0}{(f_{0}+g_{00})^2-g_{0}^{v2}}
\end{equation}

It is then clear that the resonant transmission peaks observed in Fig.2 are
associated with the zeroes of the denominator of Eq. (24). In the
strict limit $a/\lambda \rightarrow 0$, $g_{00}$ goes to zero
[as $a/\lambda \: Ln(a/\lambda)$] and resonant transmission appears at the
condition $f_0=\pm g_0^v$, which straightforward algebra shows
equivalent to the usual Fabry-Perot resonant condition
$sin(k_{0}\sqrt{\epsilon} W)=0$. When $a/\lambda$ is small but not
zero, the transmission peaks are shifted and broadened due to the
non-negligible contribution of the real and imaginary part 
of $g_{00}$, respectively. Therefore, we can say that
a subwavelength metallic slit behaves basically as a Fabry-Perot
interferometer made of high reflective plates. Following with
this analogy, it can be shown that when the slit width increases,
the modulus of the reflection coefficients at the edges of the
slit decreases, and therefore the efficiency of the system as a
interferometer becomes poorer.

The numerical results shown in Fig. 2(b) are in agreement with 
experimental data published by Yang and Sambles\cite{Sambles2}. 
However, it is worth noticing that for this case the expression obtained
in Ref. 11, where only the leading terms in $a/\lambda$ are
considered, overestimates (about 10\%) the shift of resonant peaks from the 
usual Fabry-Perot resonant condition, $\sin(k_0 \sqrt{\epsilon} W)=0$, 
showing that the asymptotic $a/\lambda \to 0$ limit is reached very slowly.

Another interesting feature observed in the subwavelength regime
is that transmittance does not depend on the angle of incidence.
This can be easily understood by looking at Eq.(11). In this
expression, the only quantity that depends on $\theta$ is $I_0$

\begin{equation}
I_{0} (\theta) = 2 \: \textrm{sinc} \left[ \pi \left( \frac{a}{\lambda} \right) \sin \theta \right]
\end{equation}

where $sinc(x)=sin(x)/x$. If $a/\lambda$ is small enough the dependence 
of this coupling on the incident angle is negligible as $sinc \approx 1$, 
independent on $\theta$.

\begin{figure}
\begin{center}
\includegraphics[width=8.5 cm]{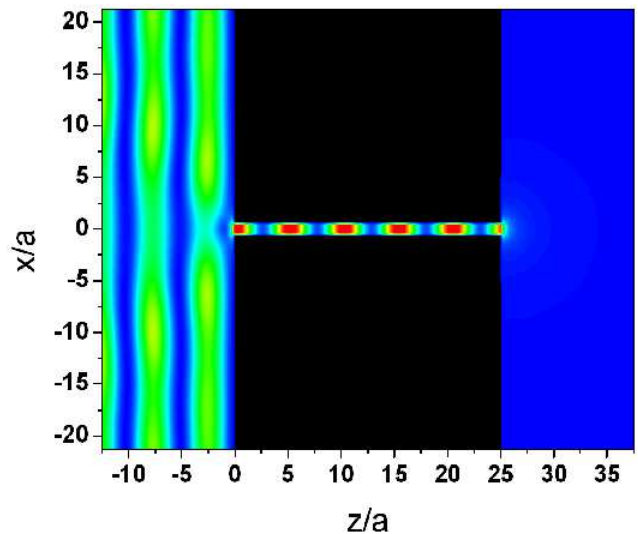}
\end{center}
\caption{Contour plot of the electric field intensity of a single metallic slit defined by
$a$=40 $\mu$m, $W$=1 mm. Incident radiation with $\lambda$=409 $\mu$m and $\theta$=0 is assumed.
Red (blue) means maximum (minimum) field intensity.}
\end{figure}

Notice that, as it was mentioned before, we are normalizing the
transmittance to the energy that reaches the slit. Therefore, 
since the total transmission in the resonant peaks for $a$=25$\mu$m 
is about three times that of $a$=75 $\mu$m, we can deduce
that \emph{the energy flux collected by the slit at resonances is
approximately the same for any value of $a$}. This extraordinary
property also implies that the electric field intensity inside the
slit satisfies a $1/a$ dependence, and therefore, huge enhancements of
the electric field intensity can be achieved just by reducing the
slit width. However, as a leaky standing wave is excited within the slit,
only a periodic set of regions will undergo the enhancement of the
electric field intensity, as is shown in the contour plot of Fig. 3, 
where we have considered a representative example with
$a$=40 $\mu$m and $W$=1 mm. Therefore, introducing a suitable
non-linear medium inside the slit would produce a periodic
modulation of the refractive index in the direction of
propagation. In other words, an one-dimensional photonic crystal is
formed by the electromagnetic radiation at the resonant
frequencies. 

It is also worth commenting that, as illustrated in Fig.2, the
linewidth of the transmission resonances decreases when the slit
width is reduced. This property allows us to measure the dielectric
constant of small amounts of material filling the slit through the variation of
the resonant wavelengths with $\epsilon$, as has been 
suggested by Yang and Sambles\cite{Sambles3,Sambles4}. In this paper, we study
the accuracy of this kind of structure to work as a 
high-sensitivity microfluidics detector. Within this context, very recently,
changes in transmission properties of photonic crystals have been
proposed as a way to identify certain very small quantities of
fluids \cite{Topolancik}. Here, we analyze a simple but
functional alternative to this approach. A small amount of a fluid
can be guided to flow within the slit and just by recording the
location of the transmission resonances, an estimation of the
dielectric constant of the material can be obtained. To prove this
proposal, Fig. 4a shows a contour plot of the transmittance as a
function of the wavelength and the dielectric constant inside the
slit. The efficiency of this system is shown in of Fig. 4b where
the same two fluids considered in Ref. 18 are assumed to fill the
slit, i.e., $\epsilon$=1.8978 (isopropanol) and $\epsilon$=2.2506
(xylene). As can be seen from this figure, the two peaks are well
separated and, in principle, with a slit with the geometrical 
parameters we have used in this simulation, this device could distinguish whether
isopropanol or xylene are filling the slit. If we were interested
in characterizing two fluids with more similar dielectric
constants, we could obtain even higher resolution in the
transmission resonances by just reducing the slit width. As
commented previously, this device also is expected to work in the optical
regime allowing then to characterize nanometric amounts of fluids.

\begin{figure}
\begin{center}
\includegraphics[width=7 cm]{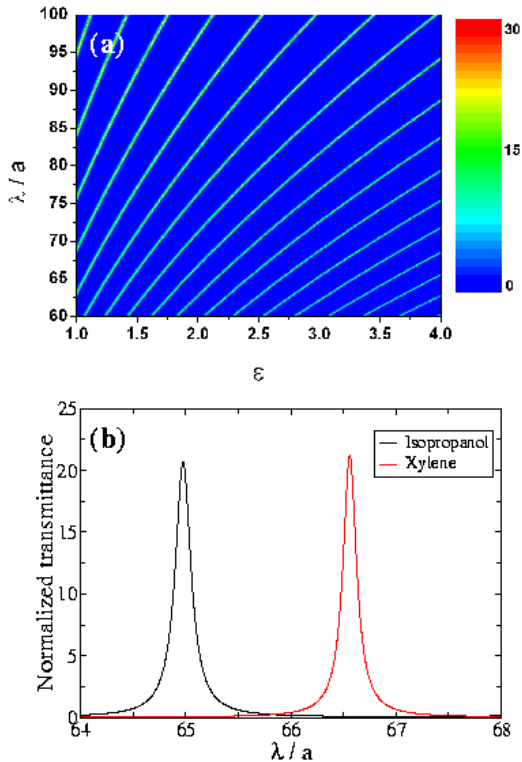}
\end{center}
\caption{(a) Contour plot of the transmittance as a function of the 
wavelength and the dielectric constant inside the slit. Geometrical
parameters as in Fig. 2(b) are assumed. (b) Comparison of the transmittance
obtained with $\epsilon$=1.8978 (black line) and $\epsilon$=2.2506 (red line).}
\end{figure}

\section{Crossover between the subwavelength regime and the geometrical-optics limit}

In the previous sections we have found no dependence of the normalized
transmittance with the angle of incidence in the extremely subwavelength 
limit. When the slit width increases, the dependence of the transmittance
on incident angle becomes important. If $a/\lambda$ is still small
enough that only the first mode ($m$=0) contributes appreciably to
the transmission, the coupling with the external plane wave is
governed by $I_0(\theta)$ [see Eq.(25)]. This is the case of the
transmittance spectrum plotted in Fig. 5(a), where we assume
$a$=1.25 mm. As shown in this figure, transmittance is maximum for
$\theta=0^{\circ}$ and decreases very slowly when $\theta$ is increased.

\begin{figure}
\begin{center}
\includegraphics[width=7 cm]{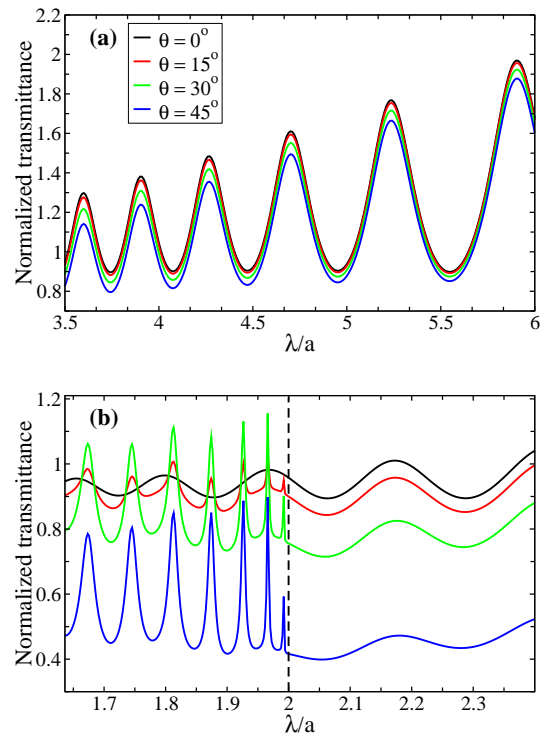}
\end{center}
\caption{Transmittance computed with W=28.2 mm and (a) a=1.25 mm and 
(b) a=2.75 mm.  Black, red, green and blue correspond to 
$\theta$ =0$^{\circ}$, 15$^{\circ}$, 30$^{\circ}$, and
45$^{\circ}$, respectively.}
\end{figure}

By increasing $a$, several eigenmodes with $m>0$  (both evanescent and
propagating) start having a non-negligible contribution to the
transmittance. The coupling of these $m>0$ modes with the external radiation
and between them is given by $I_m$ and $g_{mm'}$, respectively. An
example of this behavior is shown in Fig. 5(b), where $a$=2.75 mm.
As can be seen in this figure, when a slit mode changes its
character from evanescent to propagating (dashed line in Fig. 5(b)
shows the wavelength where $m=1$ becomes propagating,
$\lambda=2a$), resonant transmission properties change
dramatically. Interestingly, this change is not observed for
$\theta=0^{\circ}$. This is due to the fact that
$I_m(\theta=0^{\circ})=0$ for $m>0$ and $g_{0m}=0$ for $m$ odd. Then for this
particular case of $\theta=0^{\circ}$ and for this range of
wavelengths, transmission is governed by mode $m=0$. Similar
abrupt changes, at wavelengths where eigenmodes with higher values
of $m$ become propagating, have been also theoretically predicted
for the transmission spectrum of light impinging on a circular
aperture made in a perfect metal screen \cite{Roberts,deAbajo}.

\begin{figure}
\vspace{0.5 cm}
\begin{center}
\includegraphics[width=8.5 cm]{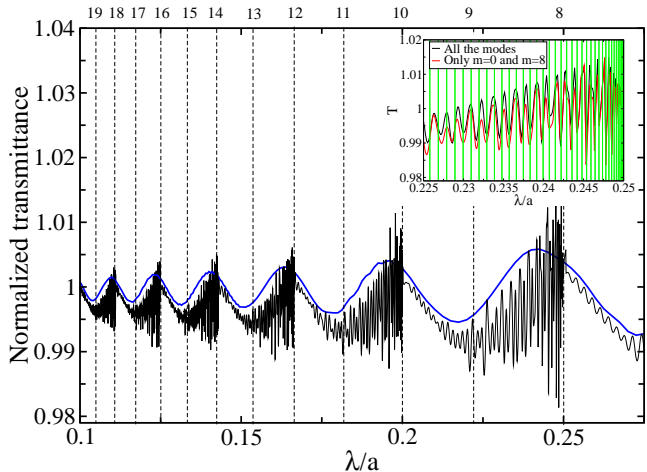}
\end{center}
\caption{Main: transmittance computed with W=28.2 mm and a=2.75 mm. Dashed lines show the wavelengths
where the corresponding mode (see labels on top) becomes propagating. Blue line corresponds to
consider only the fundamental mode. Inbox: Comparison of the
transmittance calculated taken into account all the modes (black line) and only m=0 and m=8 (red line).
Vertical green lines show the wavelengths where the Fabry-Perot condition is
fulfilled for m=8 (see text).}
\end{figure}

Now, let us focus in the interval $0.1 < \lambda/a < 0.275 $. Solid black line in Fig. 6 
shows the transmittance in this interval of wavelengths for $a$=2.75 mm. 
For the sake of simplicity, in this section we analyze only 
normal incidence. If $\theta\neq$0$^{\circ}$ similar physical 
mechanisms are present. Dashed lines show the wavelengths where the corresponding mode
changes it character from evanescent to propagating.  
First of all, notice that transmission peaks are associated to the 
emergence of propagating even modes and not to odd ones. This can be 
understood by realizing that, as commented before, for normal incidence,
odd modes are uncoupled both to the external radiation and to the
$m$=0 mode. Even modes, although uncoupled to the normal incident
radiation, have an indirect illumination through their coupling with $m$=0 mode. 
In addition, this makes that the fundamental mode approximately gives the 
overall transmission envelope (see blue line in Fig. 6).      

We have found that near the wavelength where the m-th mode (being m
a even integer) changes its character from evanescent to
propagating, i.e., at $\lambda/a=2/m$, transmission features are
mainly governed by the contribution of only the mode m=0 and the m-th
mode. In order to prove quantitatively this property, inbox of
Fig. 6 shows the transmission in the neighborhood of the
wavelength where m=8 becomes a propagating mode
($\lambda/a$=0.25). Black line in this figure shows the result
considering all the modes within the slit, while red line shows
the same calculation but now including only modes with $m$=0 and $m$=8 in
Eqs. (9) and (10). We have found that $m$=0 gives the background of the
transmittance while $m$=8 produces approximately the oscillatory behavior. 
These rapid oscillations of the spectrum can be explained by analyzing 
the $z$-component of the wavevector ($k_z$) associated to the corresponding 
eigenmode when this mode changes from evanescent to propagating. Within 
the range of wavelengths in which this change occurs, $k_z \rightarrow 0$ and 
then Fabry-Perot condition $sin(k_z\sqrt{\epsilon} W)=0$ is fulfilled for many
wavelengths inside this interval, resulting in transmission peaks
with a very small wavelength separation between them (see vertical 
green lines in inbox of Fig. 6). Finally, from Fig. 6 we can see how 
the amplitude of the transmittance oscillations decreases and tend to 1 
when $\lambda/a \to 0$, as it is expected from geometrical optics.

\section{Summary}
In this work, a formalism for analyzing the transmission
properties of a single metallic slit under general conditions of
angle of incidence and slit width has been presented. 

First of all, we have carried out a comprehensive study of the
subwavelength regime. We have discussed the appearance of
transmission resonances associated to the excitation of slit
waveguide modes. We have also discussed two possible devices based
on the huge enhancement of the EM-fields inside the slit that
accompanies the extraordinary transmission phenomenon: a
non-linear device and a micro or nano-fluidics detector.

In addition, the behavior of the transmittance when the
subwavelength limit is abandoned has been also analyzed. We have
studied in detail how the geometrical optics limit ($T=1$) is
recovered. In this regime, rapid oscillations in the transmission spectrum 
appear near the wavelength where a mode becomes propagating.

\section*{Acknowledgements}
Financial support by the Spanish MCyT under grant BES-2003-0374 and
contracts MAT2002-01534 and MAT2002-00139 is gratefully acknowledged.

\end{document}